# Managing HILP Consequences Using Dynamic Distribution System Asset Assessment

Hamidreza Sharifi Moghaddam[1], Reza Dashti[2], Abolfazl Ahmadi[3]

1. Master of Science Candidate, Iran University of Science and Technology 2. Assistant Professor, Iran University of Science and Technology 3. Assistant Professor, Iran University of Science and Technology

**Abstract**

In order to increase the resilience of distribution systems against high-impact low-probability (HILP) events, it is important to prioritize assets damaged by these events so that the lost loads, especially sensitive and important loads, can be recovered faster. For this reason, this paper discusses the prioritization of electricity supply lines for scheduling and prioritizing repair actions. To this end, the economic value of distribution system lines has been considered as a criterion representing the sensitivity of the network to hurricanes. The modeling is based on value, in which the load value, lifetime-based failure probability of the poles, fragility curve, duration of line repair by the maintenance team, and the topology factor have been considered. This is so that the significance of the demand side, the failure extent and accessibility of the lines, the importance of time, and the network configuration are considered. The results provide an order of line priority for fault resolution, in which the topology factor has a larger effect. This modeling has been tested on an IEEE 33-bus network.

## 1. Introduction

The inevitable dependence of human civilization on electricity has made this energy a critical infrastructure in modern societies. Critical infrastructures are assets, systems, and physical or virtual networks any disturbance or fault in which compromises the security, economy, and public safety and health in a country.

The performance of electrical power systems must be highly reliable under normal conditions in addition to being capable of providing high-quality power (with few outages) to customers in case of predictable events (in the field of reliability). Reliability is a well-known concept in power systems, and the reliability of a power system has been extensively studied in the previous decade [1]. Reliability in power systems indicates the capability of network components in meeting customer demands for electricity supply with acceptable power quality. The concept of reliability has also been used in industry and system engineering. In this sense, it is accompanied by statistical and probabilistic methods and describes system performance in the face of a number of predicted errors [2].

---

Author E-mail: rdashti@iust.ac.ir

In recent years, tornadoes and severe atmospheric conditions have inflicted heavy economic losses and even human casualties. Since the main goal of the power industry is the continuous and high-reliability supply of energy, improving the resilience of existing power systems against HILP natural disasters is important. The negative impact and financial costs associated with natural disasters, such as hurricanes, floods, and droughts, are expected to increase the significance of the resilience of power supply systems.

Recent HILP events show that one cannot considerably reduce the grave consequences of such events using traditional risk assessment methods. Therefore, correctly defining and improving the resilience of electricity supply infrastructure in response to HILP events is necessary more than ever to supply critical loads. Authors in [3] provide a definition of resilience and the corresponding criteria that can be used to quantitatively describe the consequences of HILP events or a group of these threats. Then, a method has been presented that can be utilized to improve the resilience of the system by optimally selecting infrastructural investments or operational activities under various limitations, including financial issues. The main focus of [4] is on how to assess power system resilience comprehensively. In this regard, a two-stage framework is proposed, and the Cost of Energy Not Supplied (CENS) is regarded as a primary criterion.

In reliability, we seek for invulnerability, and the measures taken are evaluated from an economic perspective. However, we cannot use the policy model employed in reliability for resilience because, in the latter, the probability of an event is low and its consequences are severe. Hence, we do not seek for invulnerability in resilience since it is virtually impossible and also very costly. On the other hand, there are financial limitations in distribution companies, and the need for convincing managers to invest in this field makes things more difficult. However, distribution companies readily accept crisis resolution as an important responsibility.

The resilience framework of a system can be divided into 3 basic steps:

1- The step before the event, which involves system reinforcement measures such as trimming, moving substations underground or up high, and using high-quality materials in the provision of instruments and tools. Many studies in power network reinforcement, such as [5], have focused on transmission network reinforcement considering natural hazards or terrorist attacks. Authors in [6] propose an optimal hardening strategy to increase the resilience of distribution system networks to protection against severe weather events. In this paper, various network hardening methods such as tower upgrade and vegetation management are considered. A preventive planning program suggests in [7] to increase the resilience of microgrid. According to the framework presented, before system was flooded, the microgrid is transferred to a state that encounter the least impact and stress from the flood. In order to do this, vulnerable equipment is identified first, all vulnerable equipment is removed, preventive measures are used to reduce downtime, and finally a solution is applied for preventive microgrid planning to protect the grid from flood threats.
2- The step immediately after the event, which involves restoration measures and emergency response such as emergency load shedding, special protective systems, islanding, and switching [1]. [8] focuses on the significant role of the microgrid and its challenges, and emphasizes interconnection issues. A two-step robust optimization model has been proposed in [9] to change the distribution network structure according to the load uncertainty. A restoration method for electricity distribution networks after natural disasters with the division of the distribution network to microgrids with distributed generators (DGs) has been presented in [10].

3- The final restoration, which involves evaluating the damage and required repairs, organizing repair teams and mobile transformers, and operating microgrids. Ref. [11] presents an online spatial risk analysis that can indicate risks in the process of forming in different parts of a power system under HILP events. A Severity Risk Index (RSI) that is monitored and supported online, evaluates the effects of HILP events on the resilience of the power system using the vulnerability of hurricanes on transmission networks. A time-dependent resilience criterion has been presented in [12] that examines several restoration prioritization plans. These resilience criteria have been used to evaluate the event severity reduction procedures, adaptation or reinforcement strategies in power distribution systems against HILPs. In the mentioned reference, a mixed-integer optimization problem has been employed to evaluate and select the best restoration strategy and other economic investments. This study [13] is an Agent-Based Modeling (ABM) approach to optimize the post-hurricane restoration procedure. ABM is capable of post-hurricane restoration and can test future scenarios. The authors of this paper found that parameters such as the number of outages, repair time range, and the number of humans can considerably affect the Estimated Time to Restoration (ETR). Moreover, other parameters such as the location of work and movement speed had slight effects on ETR. Ref. [14] proposes a system resilience reinforcement strategy after a HILP events according to the repair of damaged components and system operational dispatching. The repair tasks, unit output, and the switching plan are designed according to the results of damage evaluation. This reference proposes a mixed-integer optimization problem by combining the Vehicle Routing Problem (VRP) and DC load flow to reduce outage losses and rapidly store load supply resources.

Researchers use various methods, from traditional mathematical methods to artificial intelligence methods, to study the dimensions of the problem of enhancing the resilience of power distribution systems. In [15] a review of the methods for the electricity grid vulnerability analysis, pre-disaster recovery planning, and post-disaster restoration model is presented. [16] reviews AI capabilities for decision making in uncertain and high dimensional spaces in general, and their particular application is resilience enhancement problems, such as damage detection and estimation, cyber-physical anomaly detection, stochastic operation, and cyber security improvement.

The importance of time during and after an event for better managing HILP events and attempting to minimize damages to customers with respect to economic considerations is conspicuous. During reinforcement before the event and restoration actions after the event, one can neither make the system invulnerable via large expenditures nor restore all loads to the pre-event level via restoration actions. This indicates the significance of assessing damages, repairs, and restoration. This paper aims to investigate approaches for improving the crisis resolution process. Actions during a crisis include inspecting assets, monitoring the consumption status, and determining new consumption. During a crisis, emergency generation must correctly install, and unnecessary consumption must be eliminated, and the existing power must be distributed in order of consumption priority depending on the quantity of available consumption. After power is correctly managed, one must identify and prioritize for repairs the more sensitive and important consumptions. This paper's specifically target is the dynamic evaluation of the distribution system lines to prioritize the lines and repair the assets, restore the loads, and help the network to reach its initial resilience level. The modeling consider the load value, assets' failure probability and their lifetime using the fragility curve, duration of repair, and topology. These factors respectively used to consider the importance of the demand side, examine the accessibility of the lines, consider time, and emphasize the network configuration.

## 2. Proposed model

The modeling here is based on value, and the value of the lines is considered as a criterion representing the sensitivity of the network to any disaster (hurricane in this paper). The smaller the value assigned to the network lines, this sensitivity will fall to a lower degree. In other words, the lines will be assigned a higher resilience in the resilience ranking and a lower priority in the repair prioritization. This evaluation contains the major factors, which will be pointed out in the following.

### 2.1. Load value

$$v_i = L(i) \times 8760 \times LF(i) \times voll \qquad (1)$$

In the above equation, $L$ is the load consumed by the bus connected to the line $i$, $LF$ is the load factor index, and $voll$ is the value of the lost loads.

### 2.2. Failure probability of poles at a particular wind speed

The probability of pole tripping in the presence of extreme temperatures is explained in [1], however the hypothetical event in this paper is considered to be hurricane. Therefore, the failure probability of the poles is modeled with respect to changes in the wind speed. The poles have different lifetimes. This difference in lifetime manifests in the reliability-based failure probability, the tolerance threshold at different wind speeds, and the fracture limit of the poles. Using these values, one can plot the fracture curve of each pole.

$$FC(p_0, v_{th}, v_{max}) \qquad (2)$$

Where $p_0$ is the failure probability considering reliability and increases with an increase in lifetime. $v_{th}$ is the wind speed corresponding to the tolerance threshold of the pole and increases with an increase in lifetime. Moreover, $v_{max}$ represents the maximum wind speed that the pole can tolerate and decreases with an increase in lifetime.

Given the above explanations, the network poles, which have different lifetimes, are classified into k- classes or intervals in this research. The time intervals, which are the classification criterion, are determined using experimental studies of the vulnerability of the poles to previous hazards. Each interval has different $FC$s, as shown in Fig. 1.

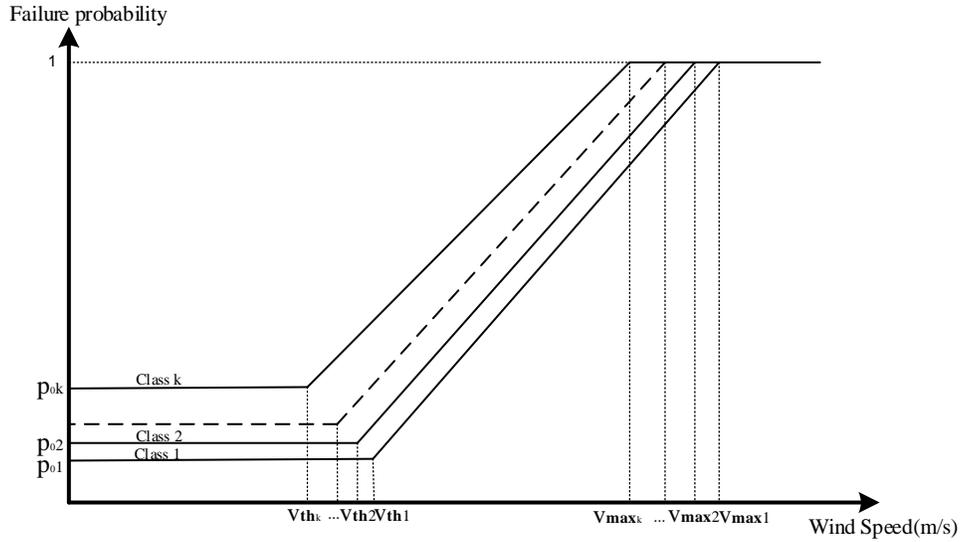

**Figure 1.** Failure probability of poles based on lifetime

In this figure, k- classes has a longer lifetime and first class has a shorter lifetime.
If the classification of assets is in the form of Table 1:

**Table 1.** Classification of assets with respect to their lifetime

|  | $R_1$<br>0 - $LFT_1$ | $R_2$<br>$LFT_1$ - $LFT_2$ | $R_3$<br>$LFT_2$ - $LFT_3$ | ... | $R_K$<br>$LFT_{K-1}$ - $LFT_K$ |
|---|---|---|---|---|---|
| $p_0$ | $p_{0_1}$ | $p_{0_2}$ | $p_{0_3}$ | ... | $p_{0_K}$ |
| $v_{th}$ | $v_{th_1}$ | $v_{th_2}$ | $v_{th_3}$ | ... | $v_{th_K}$ |
| $v_{max}$ | $v_{max_1}$ | $v_{max_2}$ | $v_{max_3}$ | ... | $v_{max_K}$ |
| $n_i$ | $n_1$ | $n_2$ | $n_3$ | ... | $n_K$ |

Where $LFT_K$ is the pole life, and $n_i$ is the number of poles in each class. Then;

$$p_{0_1} < p_{0_2} < p_{0_3} < \cdots < p_{0_K} \tag{3}$$
$$v_{th_1} < v_{th_2} < v_{th_3} < \cdots < v_{th_K} \tag{4}$$
$$v_{max_1} > v_{max_2} > v_{max_3} > \cdots > v_{max_K} \tag{5}$$

### 2.3. Number of damaged poles
In addition to the discussions in section 2.2, the wind speed ($v_{real}$) is required to obtain the failure probability of the poles with respect to changes in the wind speed. For each $R_i$ class, the graph of $FC$ will be as in Fig. 2.

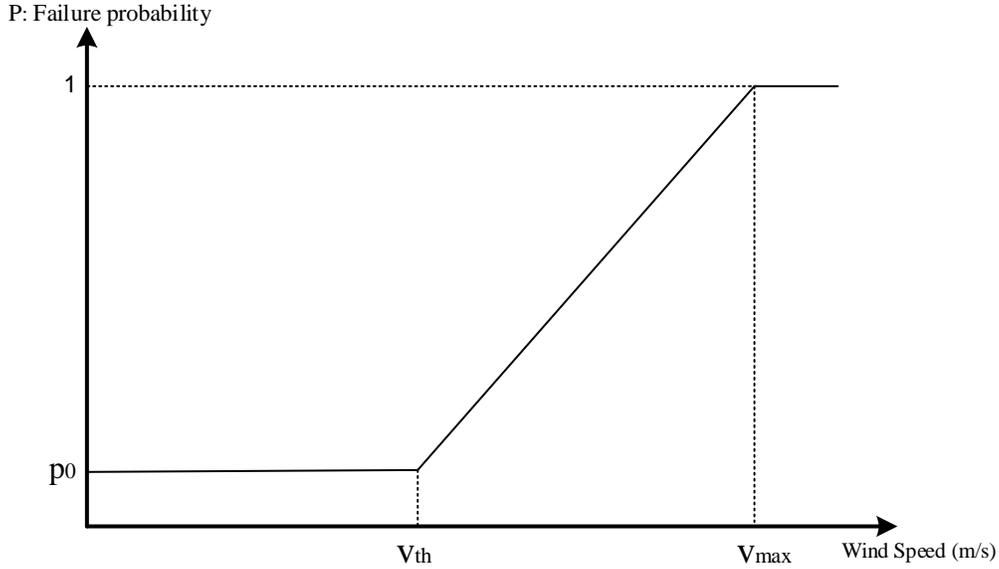

**Figure 2.** Reliability-based fragility curve of the poles

In this case, the failure probability of the poles is obtained from (6).

$$\forall i \,;\, q_{R_i} = \begin{cases} p_{0_i} & ;\, v_{real} < v_{th_i} \\ m_i(v_{real} - v_{th_i}) + p_{0_i} & ;\, v_{th_i} < v_{real} < v_{max_i} \\ 1 & ;\, v_{max_i} < v_{real} \end{cases} \qquad (6)$$

Where;

$$\forall i \,;\, m_i = \frac{1 - p_{0_i}}{v_{max_i} - v_{th_i}} \qquad (7)$$

As such, the number of damaged poles in each $R_i$ class ($b_{R_i}$) is determined using (8).

$$\forall i \,;\, b_{R_i} = q_{R_i} \cdot n_i \qquad (8)$$

### 2.4. Number of damaged poles in a line

After the event and according to the relationships examined so far, one can obtain only the number of lines requiring repair in each class (lifetime). However, there are no data regarding the number of damaged poles in each line. Hence, (9) is used to determine the number of damaged poles in each line. Given that the pole lifetime is not important during a crisis, the pole lifetime has been used in this relationship, but it has not been considered in determining the ultimate number of damaged poles.

$$bt_j = \sum_{i=1}^{4} \frac{b_{R_i}}{n_i} \cdot n_{c_{ji}} \qquad (9)$$

Where $bt_j$ is the total number of damaged poles in line $j$. Also, $b_{R_i}$ is the number of damaged poles in the class $R_i$, $n_i$ is the total number of poles in the class i, and $n_{c_{ji}}$ is the total number of poles in each line in the class i.

The above equation is used, and the time for restoring each line to the network is considered in the model, as shown in (10).

$$t_{rep_j} = bt_j \times t_{rep_{av}} \qquad (10)$$

Where $t_{rep_{av}}$ is the average time-to-repair of each pole, considered to be 4 hours.

### 2.5. Dynamic value of the load
The value of the load during a crisis is known as the dynamic value of the load. It is computed according to (11).

$$v_{dyn_i} = (L(i) \times 8760 \times LF(i) \times value).t_{rep_i} \tag{11}$$

Where $t_{rep_i}$ is duration required for the repair team to return each line to the network and is obtained from (12).

$$t_{rep_i} = \sum_{c=1}^{b} t_{rep_c} \tag{12}$$

Where $t_{rep_c}$ is the repair duration of each pole in the line, and $b$ is the number of damaged poles in the line that need repair.

### 2.6. Topology
It can be claimed that the most fundamental indicator in evaluating the lines in this paper is their position within the network. According to the modeling, the upstream lines possess also the value of all their respective downstream lines. Hence, the topology of the network is considerably effective in this sense.

### 2.7. Dynamic value of a line
The values of the lines is considered as a criterion representing the sensitivity of the network to any risk. The smaller the value assigned to the network lines, the lower the degree the sensitivity will fall. In other words, the lines will be assigned a higher resilience in the resilience ranking and a lower priority in the repair prioritization. The value of each line is obtained from the proposed formula (13).

$$v_{l_{dyn}}(i) = v_{dyn_i} + \sum_{j \in A} v_{dyn_j} \tag{13}$$

Where A is the set of lines that become unavailable in case of a disconnection in the line i and whose life and value depends on the availability of the line i. In other words, the set of lines A represents the effect of the network topology, which has the most fundamental effect on the evaluation. Since the prioritization is done during crisis and, hence, must be performed in a short time, and since the lines are compared with each other in this research, the inherent value of the line, which is related to the investment cost for building the lines, is not considered in the numerical computations that follow. The value computation model for each line is displayed in Fig. 3.

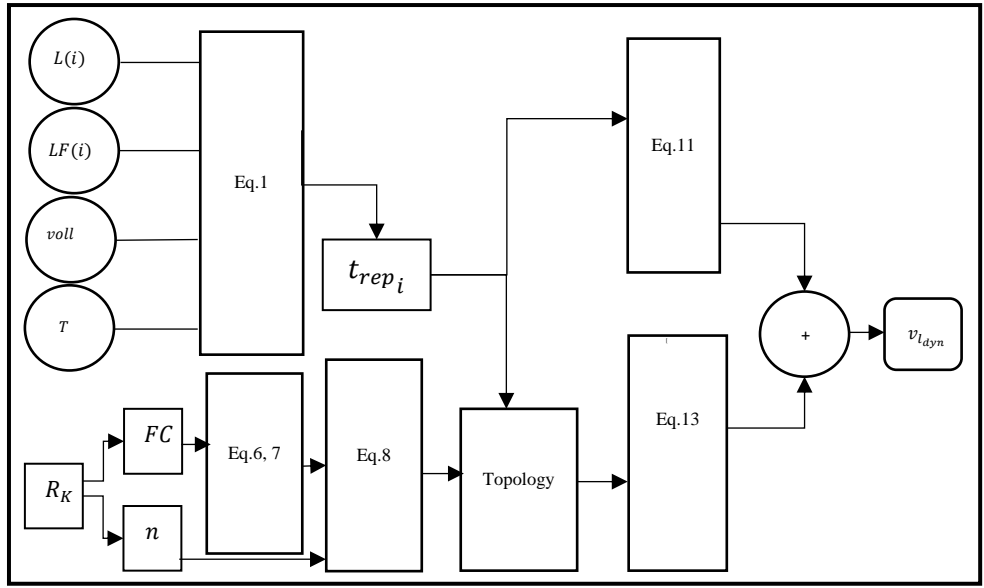

**Figure 3.** Load value computation mode

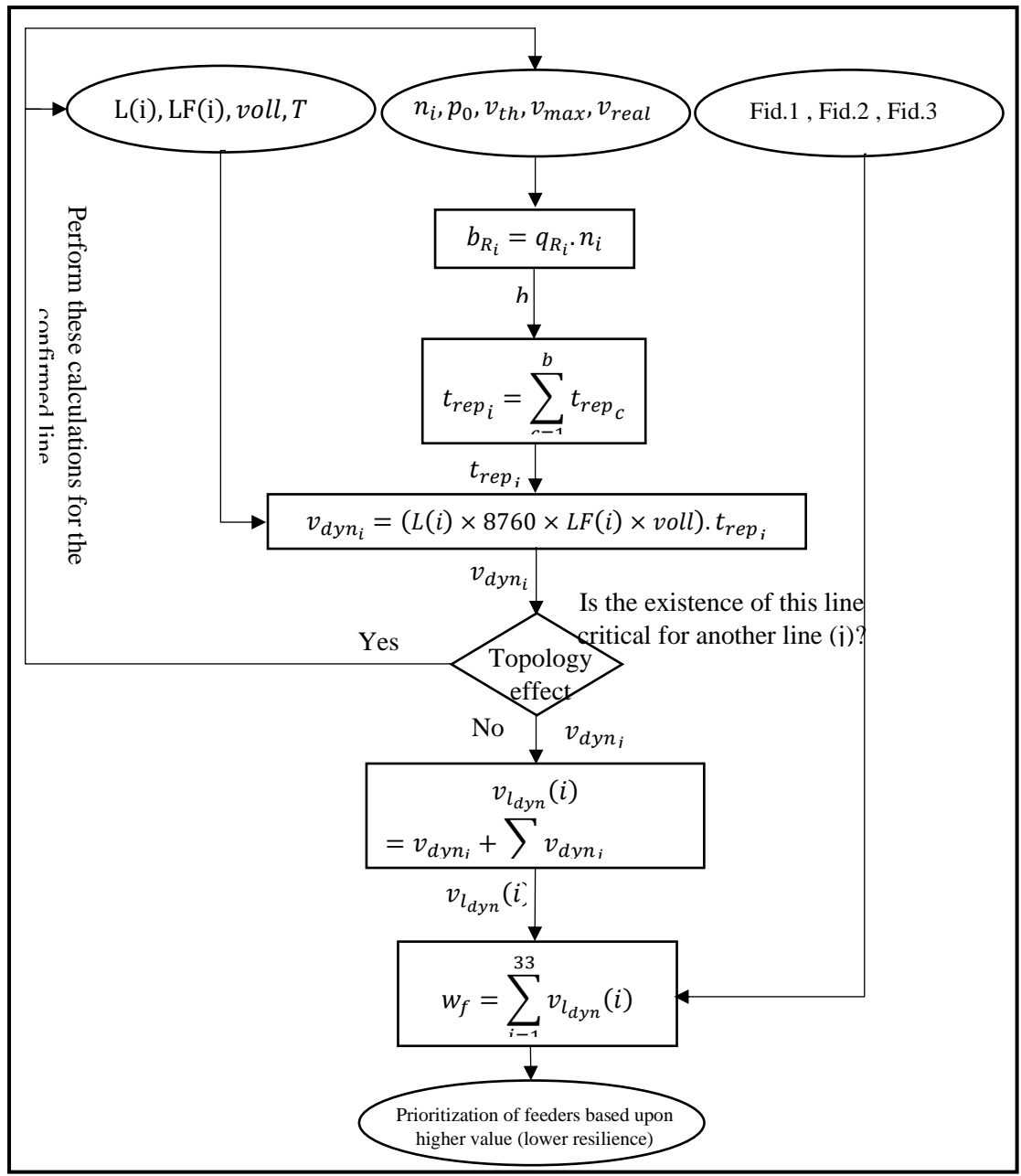

**Figure 4.** Flowchart of the proposed model

### 2.8. Dynamic value of the feeder

In order to evaluate the feeders for repair priority, the set of all evaluated lines are used, as expressed in (14). Also in this relationship, the dynamic value of the feeder is considered as a criterion for measuring the resilience of the feeder.

$$w_f = \sum_{i=1}^{33} v_{l_{dyn}}(i) \tag{14}$$

## 3. Numerical Studies

The hypothetical network of Fig. 1 is considered for implementing the proposed model. This network is composed of 3 feeders that show in Fig.5. The repairs are prioritized in two steps: Step One: Prioritization of the 3 feeders for repairs, calculated according to (12). Step Two: Evaluation of the selected feeder lines according to the main factors mentioned in section 2.

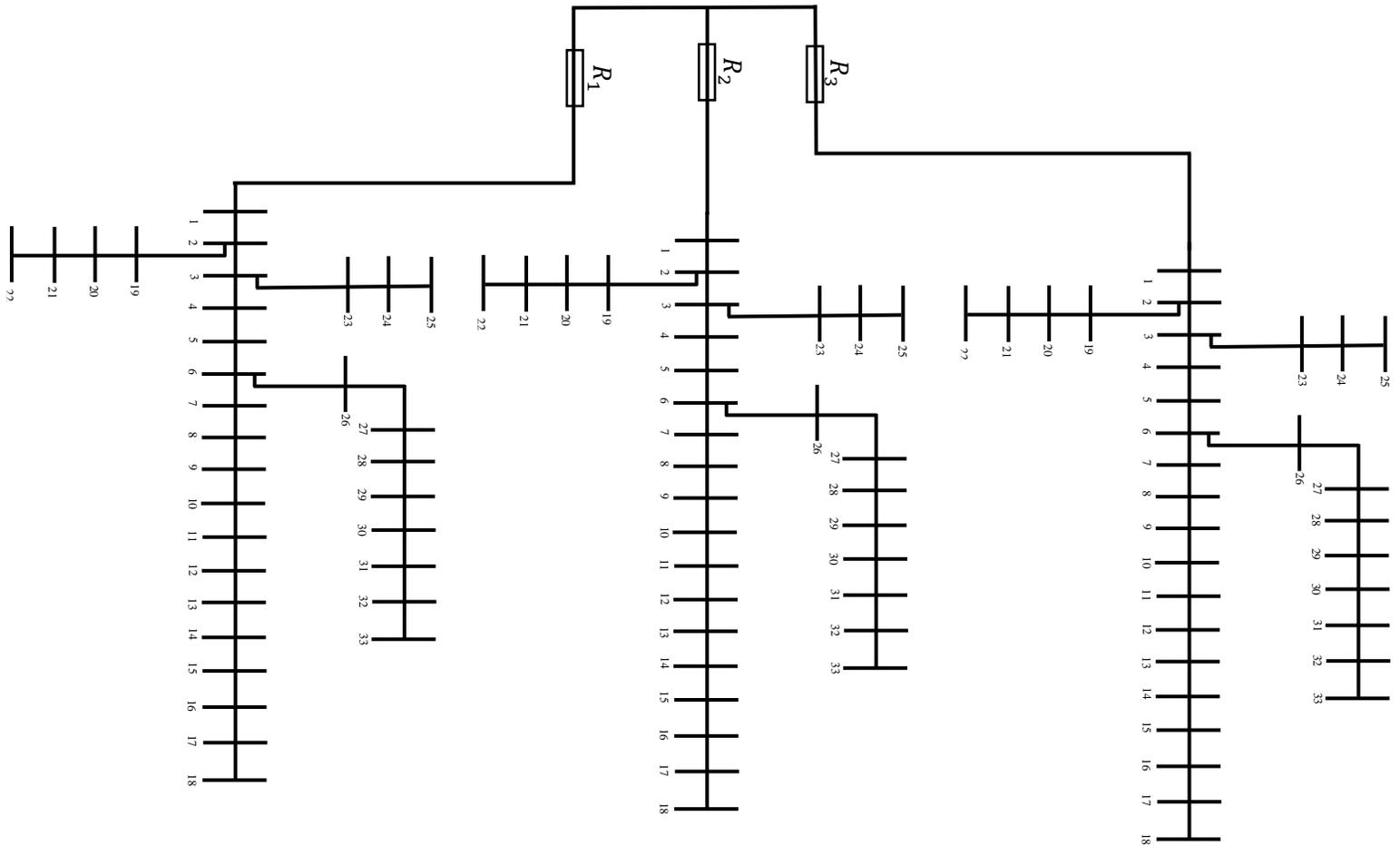

**Figure 5**. Proposed network

## 3.1. Step One

Given the data obtained from electricity distribution companies, 3 to 10 poles are installed between every 2 buses. Accordingly, an average of these numbers is considered for the 3 feeders shown in Fig. 5, such that 240, 233, and 230 poles have been considered for the first, second, and third feeder, respectively. The normal distribution function was used to distribute the poles in the 4 groups mentioned in section 2-2. The number of poles in the 3 feeders is shown in Figs. 6, 7, and 8, in which each column represents the number of poles, and the horizontal axis displays the pole lifetime.

The first feeder includes 240 poles. The pole specifications are shown in Table 2. In this table, the poles are divided into 4 groups (in terms of lifetime) the number of poles in each of which is obtained from the normal distribution. The $p_0$s, which are the failure probability considering reliability, are specified in the table for each group. Furthermore, $v_{th}$ and $v_{max}$ have been specified for each group in the table. All these data are model inputs and have been obtained from experimental data.

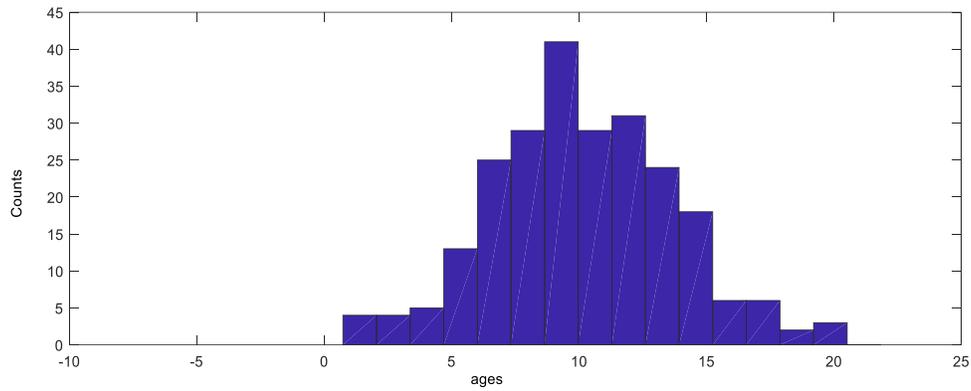

**Figure 6**. Pole distribution in the first feeder

**Table 2.** Specifications of the poles in the first feeder

| Poles | Pole lifetime (year) | Number of poles | $p_0$ | $v_{th}(m/s)$ | $v_{max}(m/s)$ |
|---|---|---|---|---|---|
| First group | 0-5 | 15 | 0.05 | 60 | 120 |
| Second group | 5-10 | 106 | 0.07 | 59.5 | 115 |
| Third group | 15-10 | 98 | 0.09 | 59 | 110 |
| Fourth group | 15-20 | 21 | 0.11 | 58 | 100 |

There are 233 and 230 poles in the second and third feeders, respectively, as shown in Tables 3 and 4.

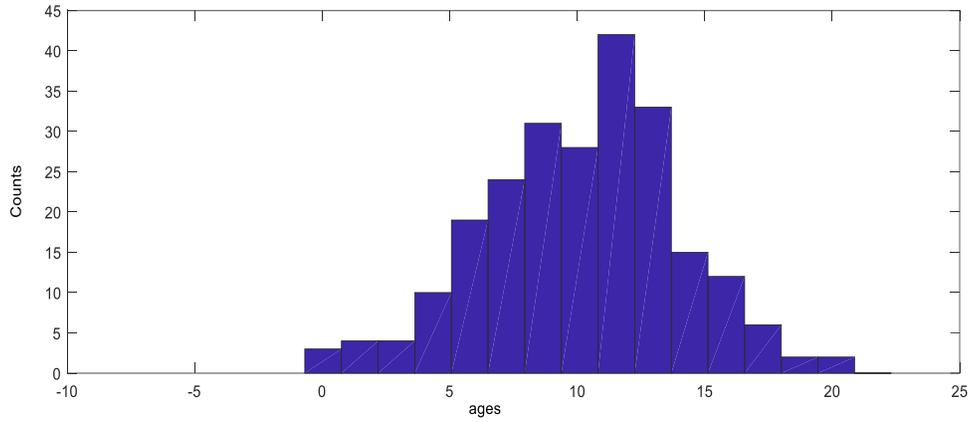

**Figure 7.** Pole distribution in the second feeder

**Table 3.** Specifications of the poles in the second feeder

| Poles | Pole lifetime (year) | Number of poles | $p_0$ | $v_{th}(m/s)$ | $v_{max}(m/s)$ |
|---|---|---|---|---|---|
| First group | 0-5 | 18 | 0.05 | 60 | 120 |
| Second group | 5-10 | 84 | 0.07 | 5.59 | 115 |
| Third group | 15-10 | 110 | 0.09 | 59 | 110 |
| Fourth group | 15-20 | 20 | 0.11 | 58 | 100 |

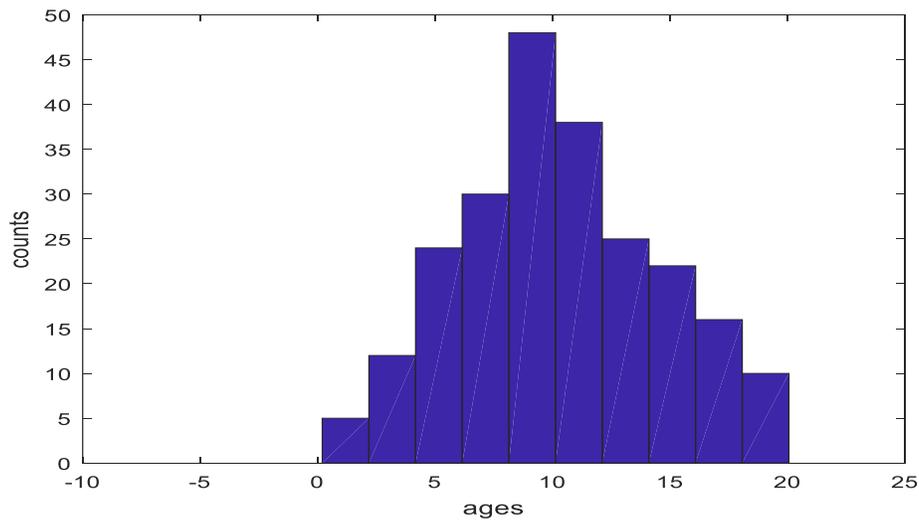

**Figure 8.** Pole distribution in the third feeder

**Table 4.** Specifications of the poles in the third feeder

| $v_{max}(m/s)$ | $v_{th}(m/s)$ | $p_0$ | Number of poles | Pole lifetime (year) | Poles |
|---|---|---|---|---|---|
| 120 | 60 | 0.05 | 17 | 5-0 | First group |
| 115 | 5.59 | 07.0 | 93 | 5-10 | Second group |
| 110 | 59 | 09.0 | 105 | 15-10 | Third group |
| 100 | 58 | 11.0 | 15 | 20-15 | Fourth group |

The poles in each feeder are divided into 4 groups. These 4 groups have different parameters. Fig. 9 displays these 4 groups.

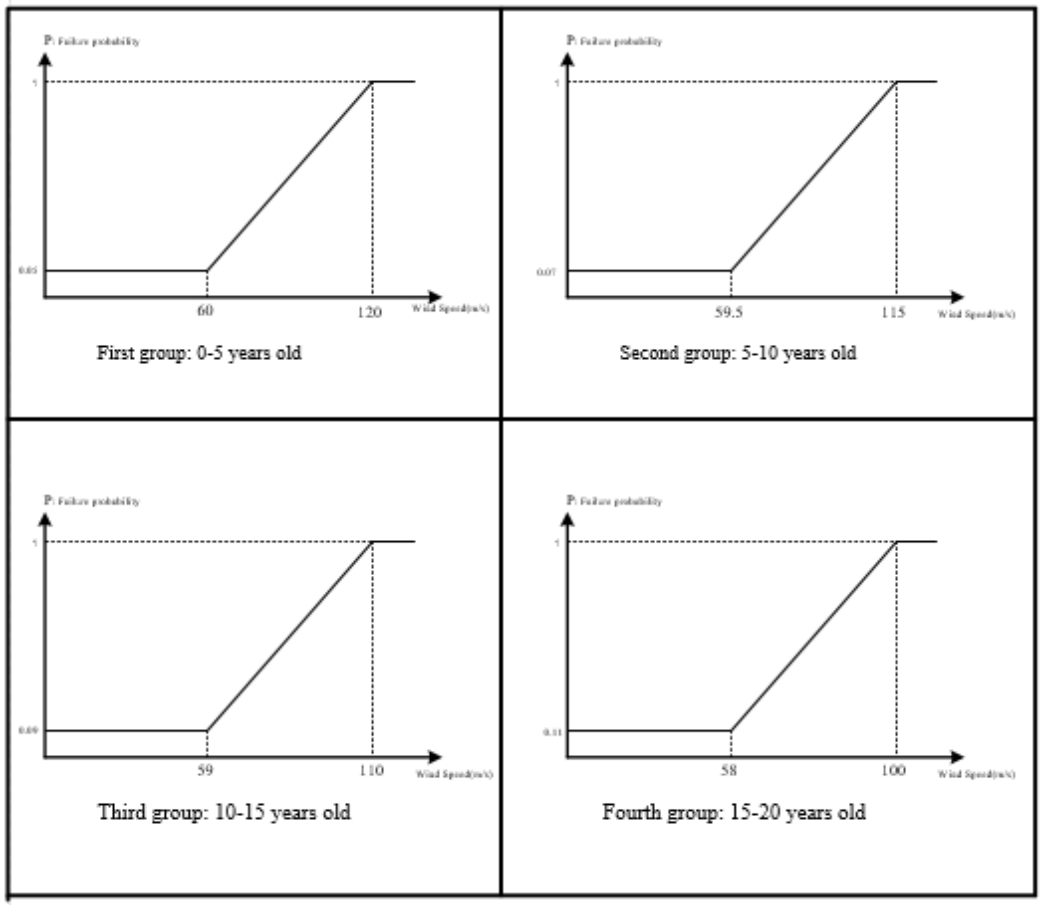

**Figure 9.** Fragility curve of poles based on the lifetime

The number of poles in each line are shown divided into groups in Table 5. This information has been considered according to experimental data. For example, there are 4 poles in the second line of the first feeder, with one pole in the third group and three poles in the fourth group.

Table 5. Number of poles in each line based on the lifetime

| line | Total fid.n1 | 0-5 | 5-10 | 10-15 | 15-20 | line | Total fid.n.2 | 0-5 | 5-10 | 10-15 | 15-20 | line | Total fid.n.3 | 0-5 | 5-10 | 10-15 | 15-20 |
|---|---|---|---|---|---|---|---|---|---|---|---|---|---|---|---|---|---|
| 1 | 4 | - | 1 | - | 3 | 1 | 4 | - | 2 | - | 2 | 1 | 4 | 1 | 3 | - | - |
| 2 | 4 | - | 1 | - | 3 | 2 | 4 | - | 1 | - | 3 | 2 | 4 | - | 1 | 2 | 1 |
| 3 | 6 | - | 2 | - | 4 | 3 | 5 | - | 2 | - | 3 | 3 | 5 | - | 3 | 2 | - |
| 4 | 7 | 1 | 3 | 2 | 1 | 4 | 7 | - | 2 | 3 | 2 | 4 | 7 | - | 3 | 4 | - |
| 5 | 6 | - | 2 | 2 | 2 | 5 | 5 | 1 | 1 | 2 | 1 | 5 | 5 | 1 | 1 | 2 | 1 |
| 6 | 6 | 1 | - | 4 | 1 | 6 | 6 | - | 1 | 3 | 2 | 6 | 5 | - | 1 | 4 | - |
| 7 | 7 | - | 2 | 4 | 1 | 7 | 7 | 1 | 1 | 4 | 1 | 7 | 7 | 1 | 2 | 4 | - |
| 8 | 7 | 1 | 3 | 2 | 1 | 8 | 7 | - | 2 | 5 | - | 8 | 7 | - | 2 | 5 | - |
| 9 | 7 | - | 3 | 4 | - | 9 | 7 | 1 | 3 | 2 | 1 | 9 | 7 | 1 | 3 | 2 | 1 |
| 10 | 8 | 1 | 3 | 4 | - | 10 | 8 | 1 | 2 | 5 | - | 10 | 8 | 1 | 3 | 4 | - |
| 11 | 7 | - | 3 | 4 | - | 11 | 6 | 1 | 2 | 3 | - | 11 | 6 | 1 | 2 | 3 | - |
| 12 | 6 | 1 | 3 | 1 | 1 | 12 | 6 | - | 1 | 4 | 1 | 12 | 6 | - | 2 | 3 | 1 |
| 13 | 8 | 1 | 2 | 5 | - | 13 | 8 | 1 | 1 | 6 | - | 13 | 8 | 1 | 1 | 6 | - |
| 14 | 8 | - | 3 | 5 | - | 14 | 7 | 1 | 1 | 5 | - | 14 | 7 | - | 2 | 5 | - |
| 15 | 7 | - | 4 | 3 | - | 15 | 7 | 1 | 3 | 2 | 1 | 15 | 7 | 1 | 3 | 3 | - |
| 16 | 8 | 1 | 4 | 3 | - | 16 | 8 | 1 | 3 | 3 | 1 | 16 | 8 | - | 3 | 4 | 1 |
| 17 | 8 | - | 5 | 3 | - | 17 | 8 | 1 | 4 | 3 | - | 17 | 8 | 1 | 4 | 3 | - |
| 18 | 9 | 1 | 4 | 4 | - | 18 | 9 | - | 4 | 5 | - | 18 | 9 | - | 5 | 4 | - |
| 19 | 10 | - | 5 | 4 | 1 | 19 | 10 | 1 | 4 | 4 | 1 | 19 | 9 | 1 | 4 | 3 | 1 |
| 20 | 8 | - | 2 | 6 | - | 20 | 8 | - | 4 | 4 | - | 20 | 8 | 1 | 4 | 2 | 1 |
| 21 | 8 | 1 | 3 | 4 | - | 21 | 7 | - | 3 | 4 | - | 21 | 7 | - | 3 | 3 | 1 |
| 22 | 7 | 1 | 6 | - | - | 22 | 7 | 1 | 5 | 1 | - | 22 | 7 | 1 | 5 | 1 | - |
| 23 | 10 | - | 4 | 5 | 1 | 23 | 10 | - | 3 | 7 | - | 23 | 10 | - | 4 | 6 | - |
| 24 | 7 | 1 | 3 | 3 | - | 24 | 7 | 1 | 2 | 3 | 1 | 24 | 7 | 1 | 2 | 3 | 1 |
| 25 | 8 | - | 3 | 5 | - | 25 | 8 | - | 3 | 5 | - | 25 | 7 | - | 3 | 3 | 1 |
| 26 | 9 | 1 | 4 | 3 | 1 | 26 | 8 | 1 | 3 | 4 | - | 26 | 8 | 1 | 3 | 4 | - |
| 27 | 8 | 1 | 4 | 3 | - | 27 | 8 | 1 | 3 | 4 | - | 27 | 8 | - | 3 | 4 | 1 |
| 28 | 6 | - | 3 | 2 | 1 | 28 | 6 | 1 | 2 | 3 | - | 28 | 6 | 1 | 2 | 3 | - |
| 29 | 7 | 1 | 3 | 3 | - | 29 | 6 | - | 2 | 4 | - | 29 | 6 | - | 2 | 3 | 1 |
| 30 | 7 | - | 2 | 5 | - | 30 | 7 | 1 | 2 | 4 | - | 30 | 7 | 1 | 2 | 3 | 1 |
| 31 | 8 | - | 3 | 5 | - | 31 | 8 | - | 2 | 6 | - | 31 | 8 | - | 2 | 6 | - |
| 32 | 7 | 1 | 6 | - | - | 32 | 7 | 1 | 5 | 1 | - | 32 | 7 | 1 | 5 | - | 1 |
| 33 | 7 | - | 7 | - | - | 33 | 7 | - | 5 | 2 | - | 33 | 7 | - | 5 | 1 | 1 |

The failure probability of the poles in the 3 feeders with respect to wind speed was obtained as shown in Fig. 10. The number of damaged poles is shown according to equation (6), (7), and (8). In this relationship, decimals are rounded to the upper integer.

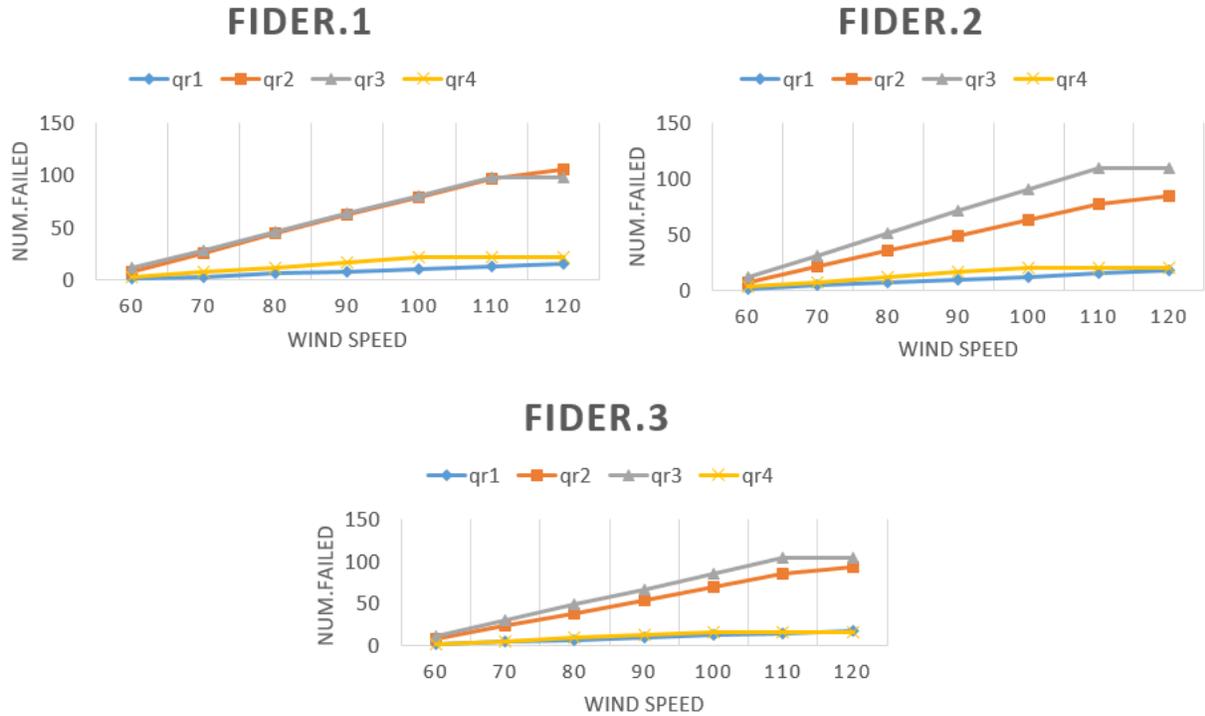

**Figure 10.** Damaged poles based on feeders and lifetime

As seen in the figure, with an increase in the wind speed, the number of damaged poles in each group and, hence, in each feeder increases. For the rest of the calculations, the wind speed is assumed to be $v_{real} = 80 \, m/s$. In this case, the number of damaged poles in each feeder based on the lifetime is as shown in Table 6.

**Table 6.** Number of damaged poles based on groups at a wind speed of $v_{real} = 80 \, m/s$

|       | $b_{R_1}$ | $b_{R_2}$ | $b_{R_3}$ |
|-------|-----------|-----------|-----------|
| $R_1$ | 6         | 7         | 6         |
| $R_2$ | 44        | 35        | 38        |
| $R_3$ | 46        | 31        | 49        |
| $R_4$ | 12        | 12        | 9         |
| Total | 108       | 105       | 102       |

In the studied network (Fig. 5), to obtain the exact location of the damaged poles in each line, the locations of the damaged poles are determined for each group using GIS. Moreover, the repair team is considered to be accommodated at the slack bus. The average repair duration for each pole is considered to be 4 hours. In addition, the maximum time for the repair team to arrive at the repair location is considered to be 1 hour and varies depending on the distance of the line from the repair team. The results are shown in Table 7, where the first number represents the damaged pole(s) that need repair.

Table 7. Number of damaged poles based on line and lifetime

| line | Total fault fid.n1 | 0-5 | 5-10 | 10-15 | 15-20 | $t_{rep_i}$ | line | Total fault fid.n2 | 0-5 | 5-10 | 10-15 | 15-20 | $t_{rep_i}$ | line | Total fault fid.n3 | 0-5 | 5-10 | 10-15 | 15-20 | $t_{rep_i}$ |
|---|---|---|---|---|---|---|---|---|---|---|---|---|---|---|---|---|---|---|---|---|
| 1 | 2/4 | - | 1/1 | - | 1/3 | 8 | 1 | 2/4 | - | 1/2 | - | 1/2 | 8 | 1 | 1/4 | 1 | 1/3 | - | - | 4 |
| 2 | 1/4 | - | 1 | - | 1/3 | 4 | 2 | 2/4 | - | 1 | - | 2/3 | 8 | 2 | 2/4 | - | 1/1 | 1/2 | 1 | 8 |
| 3 | 3/6 | - | 1/2 | - | 2/4 | 12.1 | 3 | 4/5 | - | 2/2 | - | 2/3 | 16.1 | 3 | 2/5 | - | 1/3 | 1/2 | - | 8.1 |
| 4 | 4/7 | 1 | 2/3 | 1/2 | 1/1 | 16.3 | 4 | 3/7 | - | 1/2 | 1/3 | 1/2 | 12.3 | 4 | 3/7 | - | 1/3 | 2/4 | - | 12.3 |
| 5 | 3/6 | - | 1/2 | 1/2 | 1/2 | 12.4 | 5 | 3/5 | 1/1 | 1/1 | 1/2 | 1 | 12.4 | 5 | 5/5 | 1/1 | 1/1 | 2/2 | 1/1 | 20.4 |
| 6 | 4/6 | 1/1 | - | 2/4 | 1/1 | 16.5 | 6 | 2/6 | - | 1 | 1/3 | 1/2 | 8.5 | 6 | 2/5 | - | 1 | 2/4 | - | 8.5 |
| 7 | 3/7 | - | 1/2 | 2/4 | 1 | 12.5 | 7 | 4/7 | 1 | 1/1 | 2/4 | 1/1 | 16.5 | 7 | 2/7 | 1 | 1/2 | 1/4 | - | 8.5 |
| 8 | 2/7 | 1 | 1/3 | 1/2 | 1 | 8.5 | 8 | 1/7 | - | 2 | 1/5 | - | 4.5 | 8 | 2/7 | - | 2 | 2/5 | - | 8.5 |
| 9 | 4/7 | - | 2/3 | 2/4 | - | 16.6 | 9 | 5/7 | 1/1 | 2/3 | 1/2 | 1/1 | 20.6 | 9 | 4/7 | 1/1 | 1/3 | 1/2 | 1/1 | 16.6 |
| 10 | 3/8 | 1/1 | 3 | 2/4 | - | 12.6 | 10 | 4/8 | 1 | 2/2 | 2/5 | - | 16.6 | 10 | 3/8 | 1 | 1/3 | 2/4 | - | 12.6 |
| 11 | 4/7 | - | 1/3 | 3/4 | - | 16.7 | 11 | 2/6 | 1 | 1/2 | 1/3 | - | 8.7 | 11 | 4/6 | 1/1 | 1/2 | 2/3 | - | 16.7 |
| 12 | 1/6 | 1 | 3 | 1 | 1/1 | 4.7 | 12 | 2/6 | - | 1 | 1/4 | 1/1 | 8.7 | 12 | 2/6 | - | 2 | 2/3 | 1 | 8.7 |
| 13 | 5/8 | 1/1 | 1/2 | 3/5 | - | 20.8 | 13 | 4/8 | 1/1 | 1/1 | 2/6 | - | 16.8 | 13 | 3/8 | 1 | 1 | 3/6 | - | 12.8 |
| 14 | 3/8 | - | 1/3 | 2/5 | - | 12.8 | 14 | 1/7 | 1 | 1 | 1/5 | - | 4.8 | 14 | 4/7 | - | 2/2 | 2/5 | - | 16.8 |
| 15 | 4/7 | - | 2/4 | 2/3 | - | 16.9 | 15 | 4/7 | 1/1 | 1/3 | 1/2 | 1/1 | 16.9 | 15 | 4/7 | 1/1 | 1/3 | 2/3 | - | 16.9 |
| 16 | 5/8 | 1/1 | 2/4 | 2/3 | - | 20.9 | 16 | 3/8 | 1 | 2/3 | 1/3 | 1 | 12.9 | 16 | 4/8 | - | 1/3 | 2/4 | 1/1 | 16.9 |
| 17 | 5/8 | - | 3/5 | 2/3 | - | 21 | 17 | 6/8 | 1/1 | 3/4 | 2/3 | - | 25 | 17 | 3/8 | 1 | 2/4 | 1/3 | - | 13 |
| 18 | 3/9 | 1 | 1/4 | 2/4 | - | 13 | 18 | 3/9 | - | 1/4 | 2/5 | - | 13 | 18 | 4/9 | - | 3/5 | 1/4 | - | 17 |
| 19 | 6/10 | - | 3/5 | 2/4 | 1/1 | 24.1 | 19 | 3/10 | 1 | 1/4 | 1/4 | 1/1 | 12.1 | 19 | 7/9 | 1/1 | 2/4 | 3/3 | 1/1 | 28.1 |
| 20 | 3/8 | - | 2 | 3/6 | - | 12.2 | 20 | 3/8 | - | 1/4 | 2/4 | - | 12.2 | 20 | 4/8 | 1 | 2/4 | 1/2 | 1/1 | 16.2 |
| 21 | 3/8 | 1/1 | 1/3 | 1/4 | - | 12.3 | 21 | 3/7 | - | 2/3 | 1/4 | - | 12.3 | 21 | 4/7 | - | 1/3 | 2/3 | 1/1 | 16.3 |
| 22 | 4/7 | 1/1 | 3/6 | - | - | 16.3 | 22 | 2/7 | 1 | 2/5 | 1 | - | 8.3 | 22 | 3/7 | 1 | 2/5 | 1/1 | - | 12.3 |
| 23 | 5/10 | - | 2/4 | 2/5 | 1/1 | 20.1 | 23 | 3/10 | - | 1/3 | 2/7 | - | 12.1 | 23 | 6/10 | - | 2/4 | 4/6 | - | 24.1 |
| 24 | 2/7 | 1 | 1/3 | 1/3 | - | 8.2 | 24 | 1/7 | 1 | 1/2 | 3 | 1 | 4.2 | 24 | 2/7 | 1 | 1/2 | 3 | 1/1 | 8.2 |
| 25 | 2/8 | - | 3 | 2/5 | - | 8.3 | 25 | 2/8 | - | 3 | 2/5 | - | 8.3 | 25 | 1/7 | - | 3 | 1/3 | 1 | 4.3 |
| 26 | 5/9 | 1 | 2/4 | 2/3 | 1/1 | 20.6 | 26 | 2/8 | 1/1 | 1/3 | 4 | - | 8.6 | 26 | 2/8 | 1 | 1/3 | 1/4 | - | 8.6 |
| 27 | 3/8 | 1 | 2/4 | 1/3 | - | 12.6 | 27 | 2/8 | 1 | 1/3 | 1/4 | - | 8.6 | 27 | 3/8 | - | 1/3 | 1/4 | 1/1 | 12.6 |
| 28 | 3/6 | - | 1/3 | 1/2 | 1/1 | 12.7 | 28 | 2/6 | 1/1 | 1/2 | 3 | - | 8.7 | 28 | 5/6 | 1/1 | 2/2 | 2/3 | - | 20.7 |
| 29 | 2/7 | 1 | 1/3 | 1/3 | - | 8.8 | 29 | 1/6 | - | 2 | 1/4 | - | 4.8 | 29 | 3/6 | - | 1/2 | 1/3 | 1/1 | 12.8 |
| 30 | 1/7 | - | 2 | 1/5 | - | 4.9 | 30 | 1/7 | 1 | 1/2 | 4 | - | 4.9 | 30 | 2/7 | 1 | 1/2 | 1/3 | 1 | 8.9 |
| 31 | 3/8 | - | 1/3 | 2/5 | - | 12.9 | 31 | 2/8 | - | 1/2 | 1/6 | - | 8.9 | 31 | 2/8 | - | 2 | 2/6 | - | 8.9 |
| 32 | 4/7 | 1 | 4/6 | - | - | 17 | 32 | 2/7 | 1 | 2/5 | 1 | - | 9 | 32 | 2/7 | 1 | 2/5 | - | 1 | 9 |
| 33 | 3/7 | - | 3/7 | - | - | 13 | 33 | 1/7 | - | 1/5 | 2 | - | 5 | 33 | 2/7 | - | 2/5 | 1 | 1 | 9 |

In the computations related to the load value for the feeders, all parameters except the supplied load of each feeder are considered identical. Moreover, the total load supplied by each feeder is identical. The simulation results of the 3 feeders are shown in Table 8.

Table 8. Dynamic values of the 3 feeders

| Fid.num | Fid.1 | Fid.2 | Fid.3 |
|---|---|---|---|
| $w_f$ | $1.1586 \times 10^{14}$ | $0.80588 \times 10^{14}$ | $1.0092 \times 10^{14}$ |

In the computations, the repair priorities were given to the first, third, and second feeders, in order.

### 3.2. Step Two

In the first step, which involved feeder evaluation, the repair priority was given to the first feeder. Now, given the main factors expressed in the second section and using the proposed formula (12), the value of each line in the first feeder, as the feeder with the lowest resilience, is determined in the second step.

The specifications of the actual power received by the first feeder were presented in Table 9.

**Table 9:** Specifications of the power received by the buses

| Sending bus | Receiving bus | L | Sending bus | Receiving bus | L | Sending bus | Receiving bus | L |
|---|---|---|---|---|---|---|---|---|
| 1 | 2 | 100 | 12 | 13 | 60 | 23 | 24 | 420 |
| 2 | 3 | 90 | 13 | 14 | 120 | 24 | 25 | 420 |
| 3 | 4 | 120 | 14 | 15 | 60 | 6 | 26 | 60 |
| 4 | 5 | 60 | 15 | 16 | 60 | 26 | 27 | 60 |
| 5 | 6 | 60 | 16 | 17 | 60 | 27 | 28 | 60 |
| 6 | 7 | 200 | 17 | 18 | 90 | 28 | 29 | 120 |
| 7 | 8 | 200 | 2 | 19 | 90 | 29 | 30 | 200 |
| 8 | 9 | 60 | 19 | 20 | 90 | 30 | 31 | 150 |
| 9 | 10 | 60 | 20 | 21 | 90 | 31 | 32 | 210 |
| 10 | 11 | 45 | 21 | 22 | 90 | 32 | 33 | 60 |
| 11 | 12 | 60 | 3 | 23 | 90 | - | - | - |

Other specifications required by the 33 evaluated lines are displayed in Table 10.

**Table 10.** Specifications of the network lines

| Num.l | $LF(i)$ | $voll$ | Num.l | $LF(i)$ | $voll$ | Num.l | $LF(i)$ | $voll$ |
|---|---|---|---|---|---|---|---|---|
| 1 | 0.8 | 3200 | 12 | 0.91 | 3200 | 23 | 0.91 | 3200 |
| 2 | 0.9 | 3200 | 13 | 0.8 | 3200 | 24 | 0.8 | 3200 |
| 3 | 0.85 | 3200 | 14 | 0.9 | 3200 | 25 | 0.9 | 3200 |
| 4 | 0.88 | 3200 | 15 | 0.85 | 3200 | 26 | 0.85 | 3200 |
| 5 | 0.89 | 3200 | 16 | 0.88 | 3200 | 27 | 0.88 | 3200 |
| 6 | 0.91 | 3200 | 17 | 0.89 | 3200 | 28 | 0.89 | 3200 |
| 7 | 0.8 | 3200 | 18 | 0.91 | 3200 | 29 | 0.8 | 3200 |
| 8 | 0.9 | 3200 | 19 | 0.8 | 3200 | 30 | 0.9 | 3200 |
| 9 | 0.85 | 3200 | 20 | 0.9 | 3200 | 31 | 0.85 | 3200 |
| 10 | 0.88 | 3200 | 21 | 0.85 | 3200 | 32 | 0.88 | 3200 |
| 11 | 0.89 | 3200 | 22 | 0.88 | 3200 | 33 | 0.89 | 3200 |

According to equation (11) and considering the network topology, the simulation was performed using MATLAB software, and the results are shown in order of value in Table 11.

**Table 11.** Simulation results

| Evaluation rank | Value ($\times 10^{13}$) | Line No. | Evaluation rank | Value ($\times 10^{13}$) | Line No. | Evaluation rank | Value ($\times 10^{13}$) | Line No. |
|---|---|---|---|---|---|---|---|---|
| 1 | 1.1705 | 4 | 12 | 0.3489 | 11 | 23 | 0.1830 | 29 |
| 2 | 1.1556 | 3 | 13 | 0.3346 | 19 | 24 | 0.1361 | 24 |
| 3 | 1.0746 | 6 | 14 | 0.3083 | 27 | 25 | 0.1287 | 17 |
| 4 | 0.8918 | 1 | 15 | 0.2879 | 28 | 26 | 0.1101 | 20 |
| 5 | 0.8306 | 5 | 16 | 0.2868 | 10 | 27 | 0.0894 | 12 |
| 6 | 0.5683 | 26 | 17 | 0.2501 | 8 | 28 | 0.0807 | 30 |
| 7 | 0.4416 | 2 | 18 | 0.2021 | 31 | 29 | 0.0769 | 21 |
| 8 | 0.4378 | 7 | 19 | 0.1985 | 14 | 30 | 0.0649 | 25 |
| 9 | 0.4242 | 23 | 20 | 0.1966 | 15 | 31 | 0.0590 | 22 |
| 10 | 0.4172 | 9 | 21 | 0.1927 | 16 | 32 | 0.0388 | 18 |
| 11 | 0.3807 | 13 | 22 | 0.1852 | 32 | 33 | 0.0259 | 33 |

To better understand the evaluation in prioritizing line repair in Fig. 11, priorities 1 to 11, 12 to 22, and 23 to 33 are displayed in red, orange, and green lines, respectively.

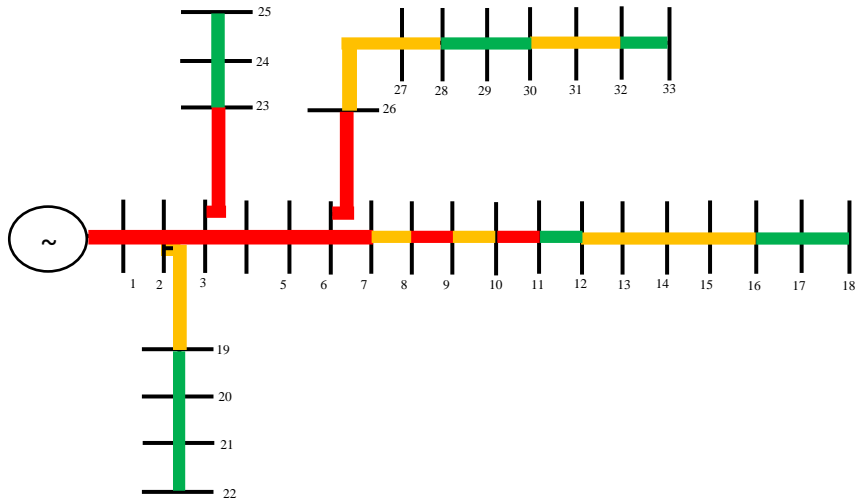

**Figure 11.** Heat map of line evaluation

Given that the number of the receiving bus represents the considered line number, the following issues are challenging: The network topology effect is clear in this heat map, such that the lines at the end of the network are at lower priorities, and the closer we get to the slack bus, the higher the value of the lines becomes. Lines 1 to 7 have the highest values due to their topological positions despite not supplying much load. Among these 7 lines, line 4 has the highest value due to the higher damage in its poles and ranks first in the evaluation ranking despite having a lower topological value compared to line 1 or 2. Another interesting issue is line 8, which ranks 17 in the evaluation ranking due to the lower vulnerability of its poles to hurricanes despite its importance due to its topology and location.

## 4. Conclusion

In this paper assets' evaluation are necessary for a fast response to HILP events in restoration from the critical condition. Generally, the lines located at the end of the network ranked lowest in the prioritization, indicating the importance of the network topology. Most of the time, relatively disperse evaluation shows the necessity of attention to repair teams, which must move fast so as to speed up the load restoration process to reach the initial resilience level. The challenge for repair planning during and after unpredictable events can be observed clearly in the obtained heat map.